\documentstyle[referee]{mn}
\input{epsf.sty}
\newif\ifAMStwofonts
\title{Relation between $M_{BH}$ and $M_{bulge}$: A simulation study}
\author[Y. N. Fu$^{1,2}$, J. H. Huang$^3$, Z. G. Deng$^4$]
       {Y. N. Fu$^{1,2}$, J. H. Huang$^3$, Z. G. Deng$^4$ \\
       $^1$Purple Mountain Observatory, Chinese Academy of
Sciences, Nanjing 210008, China\\
       $^2$National Observatories, Chinese Academy of Sciences, Beijing 100039,
China \\
       $^3$Department of Astronomy, Nanjing University, Nanjing 210093,
       China \\
       $^4$Department of Physics, Graduate School, Chinese Academy of Sciences, Beijing 100039, China}

\date{Accepted .
      Received ;
      in original form}
\pagerange{\pageref{firstpage}--\pageref{lastpage}} \pubyear{}

\begin{document}

\maketitle

\label{firstpage}

\begin{abstract}

The dynamical evolution of super star clusters (SSCs) moving in
the background of dark matter halo has been investigated as a
possible event causing the observed correlation between the mass
of galactic bulge, $M_{bulge}$, and the mass of its central black
hole, $M_{BH}$. The involved physical processes are the sinking of
SSCs due to the dynamical friction, and the stripping of SSCs on
their way to the center. Model calculations show that only sinking
of circumnuclear SSCs contribute to both the growth of the central
object and the formation of the galactic bulge at the early stage.
On the assumption of a universal density profile for the dark
matter halo, and an isothermal model for the SSCs, our simulations
have yielded the mass ratio of the central objects to the bulges
formed this way to be about a few times $10^{-4}$, less than the
observed median value for early type galaxies. It is, however,
consistent with the observed mass ratio for disk spirals, implying
that the proposed scenario might be a possible event for the
formation of bulges and central black holes of late type galaxies
and for $M_{BH} - M_{bulge}$ correlation of disk galaxies.
\end{abstract}

\begin{keywords}
black hole physics -- Galaxy: center -- Galaxy: bulge -- Galaxy:
kinematics and dynamics
\end{keywords}

\section{Introduction}

Recent observations with unprecedented high resolution have firmly
established that many galaxies, whether active or not, host
supermassive black holes (SMBHs) in their centers (Kormendy \&
Richstone 1995, Kormendy \& Gebhardt 2001). These observations
have revealed correlations of the mass of the SMBHs to the
luminosity (or mass) of their bulge (Kormendy \& Richstone 1995,
Magorrian et al. 1998, Wandel 1999), or to the luminosity-weighted
line-of-sight velocity dispersion within the effective radius with
much smaller scatter (Gebhardt et al. 2000, Merritt \& Ferrarese
2000). However, no correlation has been found between the BH
masses and the total luminosities of host galaxies (Kormendy \&
Gebhardt 2001). The median black hole (BH) mass is $0.13\%$ of the
mass of the bulge (Kormendy \& Gebhardt 2001) for early type
galaxies, or much smaller for late-type spirals (Salucci et al.
2000, Kormendy \& Gebhardt 2001, Gebhardt et al 2001).

The latest observations with X-ray Observatories provide evidence
showing that the formation of black holes is connected with
starbursts (Matsumoto et al. 2001, Fabbiano et al. 2001,
Strickland et al. 2001). Matsumoto et al. (2001) and Strickland et
al. (2001) argued that there exist intermediate-mass black holes
(MBH) in the off-nuclear compact X-ray sources in M82 and NGC3628,
respectively. The formation of SMBHs through these MBHs has been
explored soon after (Ebisuzaki et al. 2001). Haenelt \& Kauffmann
(1999) have also studied the relation between SMBHs and the
formation of galaxies. On the other hand, a beam model has been
proposed (King et al. 2001) to ease the difficulties that the
unbeamed models run into, where MBHs are required.

The correlation between the mass of SMBHs, $M_{BH}$, and the mass
of the bulge, $M_{bulge}$, implies that the growth of the central
BH and the formation of the bulge are probably caused by a same
physical process, or event (e.g. Kormendy \& Gebhardt 2001). This
"event", however, has not been clarified yet.

In this paper, we report our study on the dynamical evolution of
circumnuclear SSCs moving in the dark matter halo as a possible
"event" for the observed relation of $M_{BH}$ to $M_{bulge}$. We
find that the sinking of the SSCs, along with the tidal stripping
on their way to the center, could contribute to simultaneous
growth of $M_{BH}$ and $M_{bulge}$. Based on a set of reasonable
parameters, the mass ratio $M_{BH}/M_{bulge}$ derived from our
numerical simulations coincides with the observed values of disk
spirals.

\section{Models}

\subsection{SSC model}
Recent observations of starburst galaxies with high resolution
have revealed that there are many compact, young and very luminous
SSCs in the central regions of galaxies (Shaya et al. 1994, Surace
et al. 1998, Surace \& Sanders 1999, Whitmore et al. 1999,
Scoville et al. 2000, de Grijs et al. 2001). Numerical simulations
have also shown that interaction or merger among galaxies can
trigger strong starbursts around the nuclear regions. While
galaxies are merging or interacting, some gas components could
lose their angular momentum and fall into the central region
(Mihos \& Hernquist 1996, Barnes \& Hernquist 1996). Then, rather
high pressure of the warm interstellar gas could induce global
collapse of giant molecular clouds and thus forming circumnuclear
SSCs (Jog \& Solomon 1992, Harris \& Pudritz 1994).

On the one hand, there exist some observational facts about SSCs,
but not many conclusive results about their dynamical properties
are reached so far. On the other hand, the SSCs are believed to be
the progenitors of present-day globular clusters (GC) (e.g. Smith
\& Galagher 1999, Origlia et al. 2001). It would then be
reasonable to model SSC by using both clues mentioned above. We
assume that SSCs have a similar mass spectrum as the initial GC
mass function but with larger mean value. According to Vesperini
(2000,2001), we take the following log-normal mass function for
SSCs (SSCMF), with its mean at $5 \times 10^6 M_{\odot} $
following the investigations on SSCs (e.g. Surace \& Sanders 1999,
Origlia et al. 2001, Surace et al. 1998, Cen 2001, Meylan 2001),
\begin{equation}
\label{SSCmassspectrum} Log_{10}(M) \sim  N( \exp(mean)=5 \times
10^6, variance=0.08)
\end{equation}
With this SSCMF, we generate randomly 100 sets of SSCs, each
containing 100 SSCs (for taking 100 SSCs, see Combes 2001,
Fellhauer 2001, Whitmore et al. 1999).

For simplicity, an SSC is modeled as a truncated isothermal sphere
with three parameters: the central density ($\rho_c$), the
velocity dispersion ($\sigma$) and the initial truncated radius
($R_0$). Taking a lower and an upper limits of $\rho_c$ to be $5.3
\times 10^3 M_{\odot}/pc^3$ and $3.4 \times 10^4 M_{\odot}/pc^3$
from observed ones (Larsen et al. 2001, Campbell et al. 1992), we
assume a linear function of cluster mass, $M$, for $\rho_c(M)$.
And with $R_0$ taken to be the local tidal radius, $\sigma$ can be
derived from $M$ and $\rho_c$. The resulted $\sigma$ also
increases with $M$.

The initial distribution and evolution of SSCs in a starburst
galaxy is not clear, though it is very important for the problem
we are investigating. In our simulation, all of the 100
circumnuclear SSCs in each set are initially placed at a distance
of $1kpc$ from galactic center, typical locations for
circumnuclear SSCs, and assigned with local circular speed. As
will be discussed in the last section, we also tried larger
distance and found that more distant SSCs are generally
irresponsible for the formation and early growth of bulge and
central compact objects.

\subsection{Background}

As we are considering the formation process of bulge and central
massive objects, the dark matter dominates not only globally, but
also in the central region at early stage. Therefore, only dark
matter halo is considered initially in our simulation. For the
dark halo, we assume the universal density profile (Navarro, Frenk
\& White 1997), which can be written as (e.g., Binney et al.,
1998)
\begin{equation}
\label{NFWMassdensity} \rho_h(r)=\frac{M_{0h}}{r(a_h+r)^2}
\end{equation}
where $r$ is the distance from the halo center, and, $a_h$  and
$M_{0h}$ are two parameters. The value of $a_{h}$ is connected
with the extension of the halo and it may change with time.
However, how it changes is not well quantified. So, we take two
rather different values of $a_{h}$ for comparison. Later, we will
see that the results are not too much different, and so,
acceptable. Besides, since we are concerned with the formation and
the early growth phases of galactic bulge and its massive central
objects, of which the time duration is taken to be $1Gyr$, $a_h$
might be well approximated as time-independent. Following El-Zant
et al. (2001), we take $a_h=6kpc$ and $10kpc$, and, $M_{0h}$ is
derived from $a_{h}$ with the condition that there are $10^{12}
M_{\odot}$ interior to $200kpc$.

Our test calculations show that the first fallen SSCs, the massive
ones in each set of SSCs, can contribute about $2 \times 10^4
M_{\odot}$ to the central globe of radius $1pc$, while in the same
region there are dark matter of less than $10^3 M_{\odot}$. In the
meantime, the mass in the innermost region (interior to a few
$10pcs$ from the halo center) can be significantly changed by the
fallen SSCs, although outside about $200pc$s remains dark halo
dominating. The background variation of this kind has substantial
effect on the further mass contribution to the central region from
stripped SSCs. In order to account for the effect of fallen SSC
mass, which is characterized by a steeper mass density profile
near the center than the universal one, a truncated singular
isothermal sphere is added after the most massive SSC has fallen.

\subsection{Dynamical friction}
In the following,  $M$ and $\vec{V}_M$ (with $V_M=|\vec{V}_M|$)
denote, respectively, the mass and velocity of cluster
experiencing the dynamical friction. Assuming a Maxwellian
velocity distribution with dispersion $\sigma_{background}$ of
background matter, composed of particles with mass much smaller
than $M$, the dynamical friction formula writes (e.g., Binney et
al, 1987)
\begin{equation}
\label{Chdf} d\vec{V}_M/dt=- \frac{2 \pi \log(1+\Lambda^2) G^2 M
\rho}{V_M^3}[ {\rm erf} (X)-\frac{2X}{\sqrt{\pi}} \exp(-X^2)]
\vec{V}_M
\end{equation}
where {\rm erf} is the error function, and,
\begin{equation}
\label{quantiesinChdf} \left\{\begin{array}{l}
\Lambda=\frac{b_{max} V_{typ}^2}{GM} \\ \\
X=\frac{V_M}{\sqrt2 \sigma_{background}}
\end{array}\right.
\end{equation}

The quantity $b_{max}$ is the so-called maximum impact parameter
and $V_{typ}$ a kind of typical module of the relative velocity
between $M$ and a background particle. Neither $b_{max}$ nor
$V_{typ}$ is precisely defined. Fortunately, uncertainty in either
quantity causes no significant difference in the resulted values
of the dynamical friction. Following Binney et al. (1987), we use
$b_{max} \equiv 2kpc$ and take $V_{typ} \equiv V_M$. The velocity
dispersion $\sigma_{background}(r)$ can be roughly estimated from
the Jeans equation.

\subsection{Stripping}
It is assumed that the stellar mass outside a sphere, the radius
of which is denoted as $R_t$, approximating instant Hill stable
region around the SSC center, will be stripped. Since the
stripping is processed continuously as $r$ (the distance between
halo and SSC centers) decreases, only a thin outer layer is to be
stripped at a time. Therefore, in an average sense, the stars
stripped when the SSC goes from $r$ to $r-dr$ are contributed to a
region radially bounded by $r+R_t(r)$ and $r-dr-R_t(r-dr)$. As a
first-order approximation, the mass of the stripped stars are
considered to be, at some later epoch, uniformly distributed in
the shell bounded by $r+R_t(r)$ and $r-dr-R_t(r-dr)$. By summing
up all of the stellar mass stripped at various $r$s, the stripped
stellar mass distribution can be derived. The SSC's mass
contributed to the galactic center (taken as a globe with radius
$1pc$) is just the remaining mass of the SSC when $r+R_t (r) \leq
1pc$ plus the previously stripped mass inside the above-mentioned
$1pc$ globe.

Obviously, if a massive single object is embedded in the center of
SSC, stripping cannot be proceeded further when only this object
is left. X-ray observations discovered a lot of the so-called
super-Eddington sources associated with SSCs (Matsumoto et al.
2001, Strikland et al. 2001). However, whether there are MBHs,
ranges from several hundreds to about one thousand solar mass
(Ebisuzaki et al. 2001), or the observations are only due to beam
effect (King et al. 2001) is still not clarified. If massive black
holes do form in SSCs, they would most likely be at the center of
the SSCs. Therefore, in our simulations, we consider two extreme
cases: the stripping is not allowed when the mass of the stripped
SSC is less than $1 M_{\odot}$ and $1000M_{\odot}$, respectively.

\section{Results}
\begin{figure}
{\epsfxsize= 7.5 cm \epsfbox{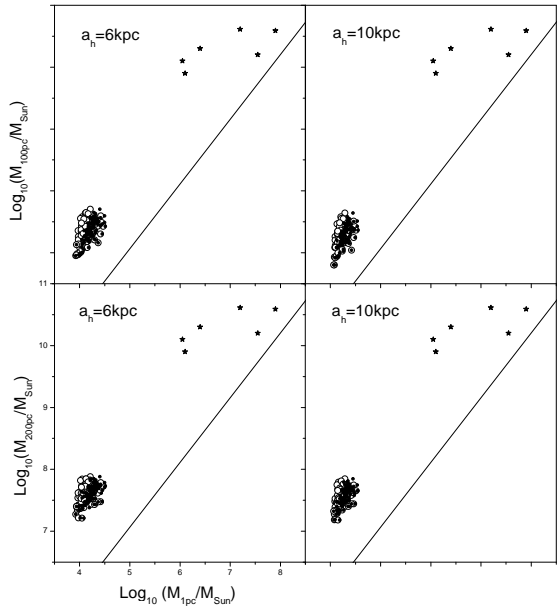}} %\end{figure*}
\vspace{-2.5cm} \caption{$M_{BH}$ vs $M_{bulge}$. Legend: solid
line $-$  results for early type galaxies; $\star\ - $
observational results for Sb-Im galaxies; open and filled circles
$- $ simulation results with and without non-striped compact cores
(or BHs) embedded in the SSCs, respectively.}
\end{figure}
The results about the relation between SSCs' mass contributions to
$M_{bulge}$ and $M_{BH}$ are summarized in  Fig.1, where
$M_{1pc}$, $M_{100pc}$ and $M_{200pc}$ are the mass contributions
of the SSCs to the globes of radii $1pc$, $100pc$ and $200pc$,
respectively. The observational results for early-type galaxies
(Kormendy \& Gebhardt 2001) and six Sb-Im galaxies (Salucci et al,
2000) are also shown. In our simulations,  94 - 96 percent of the
SSC sets, each representing SSCs in a single galaxy, can
contribute mass to the inner $1pc$ globe.  Here we take two
values, $100pc$ and $200pc$, for the radius of bulge. As shown in
Fig.1, no substantial differences between these two cases. Noting
further that the stripped SSC mass interior to $100pc$ is
generally no more a small quantity in comparison with the dark
halo mass in the same region, to take $100pc$ as the radius of a
bulge (or pseudo-bulge) form this way might be appropriate.

By comparing the left and right panels of Fig.1, one finds that
$M_{bulge}-M_{BH}$ relation has no fundamental disparity for
different values of $a_h$. This implies that similar results could
be obtained for time-dependent $a_h$, regardless how it changes
with time. Besides, the results for SSCs with and without central
black holes are also similar to each other. This is because the
mass of the formed central object of galaxy, much larger than that
of the assumed SSC's central black hole ($10^3M_{\odot}$), is
mainly comes from the most massive SSC. As a result, the obtained
$M_{bulge}-M_{BH}$ relation might be valid for generic very-late
type disk galaxies, provided that the bulge-BH do form in this
way.

As can be seen from Fig.1, $M_{BH}/M_{bulge}$ in our simulation is
smaller than that of the observed one for early type galaxies.
And, as already stated, strong tidal force of galactic central
object previously formed from the most massive SSC prohibits the
non-single objects sinking into the $1pc$ globe. This implies that
the bulge mass will be increased faster afterwards, and so, even
smaller value of $M_{BH}/M_{bulge}$ is expected at some later
stage. Our results, however, consistent with the observational
ones for disk spirals (Salucci et al. 2000, Kormendy \& Gebhardt
2001, Gebhardt et al. 2001), implying that the proposed scenario
of $M_{BH} - M_{bulge}$ formation might be valid for very-late
type disk galaxies. Possibly, the $M_{BH} - M_{bulge}$ correlation
for galaxies of various types of Hubble sequences might not be
linear, which is a possibility discussed in the case of M33
(Gebhardt et al. 2001). To detect the BH masses of less than $10^6
M_{\odot}$ in galaxies would be crucial for deliberating the
linearity of the $M_{BH} - M_{bulge}$ correlation.

\section{Discussion}
In our simulation, we have made some assumptions either due to
lack of knowledge or for simplifying the simulations.

An important assumption is that the background is composed of dark
matter only, though we have considered the background variation
later. Indeed, this assumption is compatible with what we
investigate in this paper $-$ the formation of bulges and the
growth of central black holes at the early stage. In accordance
with this, the SSCs we have studied are circumnuclear ones
assumably originated from the mergers of very late-type galaxies.
These galaxies are sources with disk components only. In this case
the dark matter of a few times $10^9 M_{\odot}$ dominates over
luminous systems inside $1~kpc$ from the center. As a first step
of our investigations, it would be reasonable to make such an
assumption. On the other hand, it will be interesting to see what
the theoretical correlation will be if more components are
assumed, e.g. disks and a pseudo-bulge, in the background for our
simulations. In other words, the logical, next step of our
investigations is to study a way for the first formed bulge and
the central BH to grow further, and to see what the ratio of
$M_{BH}/M_{bulge}$ will be. Probably, a new merger is needed. That
is, a new merger occurs between two disk galaxies with small
central BHs and pseudo-bulges. A study of this kind is under our
consideration.

Another point we have not considered in our simulations is the
distance distribution of SSCs, on an average of $1kpc$ for
circumnuclear SSCs. SSCs at different distance would make
different contribution to both bulge and central object masses.
M33 might be an example in favor of considering distance
distribution for SSCs. A compact star cluster is observed to
locate at its nucleus (e.g. Gebhardt et al. 2001), which is more
like a GC based on its dynamical parameters (Kormendy \& McClure
1993). The formation of this GC might find its way in the scenario
proposed in this paper.

Besides, in our simplified treatments, we include no effects of
non-spherical SSCs mass distribution, of background rotation, and
of others. We'll cooperate these effects in our future studies.

The survival of SSCs over Gyrs is a key question for the scenario
presented in this paper, which is also crucial for the hypothesis
where the SSCs are progenitors of present-day GCs. This question
has been analyzed by Origlia et al.(2001) for one SSC in NGC 1569,
NGC1569-A1. They found that this SSC has a standard Salpeter
initial mass function with no truncation at lower mass limit,
implying that it can evolve into a system similar to the
present-day GCs. Forbes, et al.(2001) have claimed recently that
the formation of a bulge/spheroidal stellar system is accompanied
by the formation of metal-rich GCs (red GCs). Adopting the
scenario of SSCs as the progenitors of GCs, our proposed processes
for the formation of bulge, i.e. through sinking of SSCs, along
with the tidal stripping, thereafter might be a way for their
claim.

We have also performed numerical simulations for the SSCs located
far away from the center, without considering the effect of tidal
stripping. The results show that these SSCs with masses of $10^6 -
10^7 M_{\odot}$ will stay in the external regions. Cen (2001)
proposed that the external SSCs are formed from gas-rich
sub-galactic halos triggered by the reionization of the universe.
These young stellar systems with masses of $10^3 - 10^6 M_{\odot}$
are suggested as progenitors of the present-day halo GCs (blue
GCs). Our study on the external SSCs supports Cen's suggestion in
the sense that they will stay well outside the galactic central
regions.

\section*{Acknowledgments}

The authors are grateful to Drs X.P. Wu, X.Y. Xia  and C.G. Shu.
for their valuable discussions. This work is supported by NKBRSF
G19990754, G19990756 and G19990752, and NSFC.

\end{document}